# A Behavior-Based Ontology for Supporting Automated Assessment of Interactive Systems

Thiago Rocha Silva
ICS-IRIT, Université Paul Sabatier
Toulouse, France
rocha@irit.fr

Jean-Luc Hak
ICS-IRIT, Université Paul Sabatier
Toulouse, France
jean-luc.hak@irit.fr

Marco Winckler
ICS-IRIT, Université Paul Sabatier
Toulouse, France
winckler@irit.fr

*Abstract* — Nowadays many software development frameworks implement Behavior-Driven Development (BDD) as a mean of automating the test of interactive systems under construction. Automated testing helps to simulate user's action on the User Interface and therefore check if the system behaves properly and in accordance to Scenarios that describe functional requirements. However, most of tools supporting BDD requires that tests should be written using low-level events and components that only exist when the system is already implemented. As a consequence of such low-level of abstraction, BDD tests can hardly be reused with diverse artifacts and with versions of the system. To address this problem, this paper proposes to raise the abstraction level by the means of a behavior-based ontology that is aimed at supporting test automation. The paper presents an ontology and an ontology-based approach for automating the test of functional requirements of interactive systems. With the help of a case study for the flight tickets e-commerce domain, we demonstrate how tests written using our ontology can be used to assess functional requirements using different artifacts, from low-fidelity to full-fledged UI Prototypes.

## I. Introduction

Assessing interactive systems is an activity that requires a considerable amount of efforts from development teams because it implies to assess systems features with respect to the many possible user entries and system outputs that might occur when a user is interacting with the system. Moreover, the system behavior should pass acceptance testing, which is aimed to determine if the user's point of view about a feature is in accordance with the requirements previously specified. Thus, the automation of tests for assessing the system behaviors becomes a complex task that requires the use of frameworks to simulate the user's actions when interacting with the system.

In recent years there is an increasing interests both from academic and industrial communities in Behavior Driven Development (BDD) [1][2][3] for supporting acceptance testing. One of the strengths of BDD is to support the specification of requirements in a comprehensive natural language format specification, the so-called User Stories [4]. This format allows specify executable requirements, which means we are able to test requirements specification directly, conducting to a "live" documentation and making easier to the clients to set their final acceptance tests. It guides the system development and brings the opportunity to test Scenarios directly on the User Interface (UI) with the aid of external frameworks for different platforms.

During the last seven years, we have been involved in the development of e-Government applications where we have observed certain patterns of low-level behaviors that are recurrent when writing BDD Scenarios for testing functional requirements with the UI. Besides that, we could also observe that User Stories specified in natural language often contain semantic inconsistencies. For example, it is not rare to find Scenarios that specify an action such as a selection to be made in semantically inconsistent widgets such as a Text Field. These observations motivated us to investigate the use of a formal ontology for describing pre-defined behaviors that could be used to specify Scenarios.

In this paper, we introduce the ontological model for supporting the description of Scenarios and how such Scenarios can help the automated assessment of interactive systems. The ontology was developed based on the BDD principles, describing user's behaviors when interacting with UI elements in a Scenario-based approach. Results of the ontology validation are presented by demonstration of its correctness through a consistency checking. In addition, we describe an exploratory case study that has been conducted for the flight tickets e-commerce domain. In this study, we have used our ontology-based tools to support the assessment of evolutionary Prototypes and Final User Interfaces. In the following sections, we discuss the foundations for this work, how we have built the ontological model to support the automated assessment of interactive systems, followed by its validation. We conclude with a discussion and future works.

## II. Foundations

### A. Computational Ontologies and Related Works

Computational ontologies are a mean to formally model the structure of a system, i.e., the relevant entities and relations that emerge from its observation, and which are useful for a purpose [5]. Computational ontologies come to play in this work as a means to formalize the vocabulary and the concepts used in User Stories, Scenarios and user's behaviors during the development process of interactive systems. Without a common agreement on the concepts and terms used it would be difficult to support the assessment of user requirements. Some approaches have tried to define languages or at least a common vocabulary for specifying UIs for interactive systems. Useful abstractions for describing interactive systems include the components that compose the presentation of a User Interface and the dialog parts that describe the system behavior.

The Cameleon Framework [6] treats the presentation and the dialog in three levels of abstractions: Abstract, Concrete and Final User Interfaces. The idea is that as abstract user interface component (such as a Container) could be refined to a more concrete representation (such as a Window) that will ultimately feature a final implementation in a target platform (e.g. MacOS or Windows). User Interface (UI) specifications include more or less details according to the level of abstraction. The UsiXML (USer Interface eXtensible Markup Language) [7] implements the principles of the Cameleon framework in a XML-compliant markup language featuring many dialects for treating Character User Interfaces (CUIs), Graphical User Interfaces (GUIs), Auditory User Interfaces, and Multimodal User Interfaces. UsiXML is a declarative language that captures the essence of User Interface components. At a highest level of abstraction, UsiXML describes concepts of widgets, controls, containers, modalities and interaction techniques. The dialog component of UsiXML is a state machine and it describes concepts of states, conditions, transitions and actions. Using a dedicated notation called SWC (StateWebCharts) [9], some authors [8] have demonstrated that it is possible to describe the system behavior at different levels of abstraction using UsiXML [8].

A glossary of recurrent terms for presentation components has published by the W3C in the MBUI (Model-based User Interface domain) [10]. It was intended to capture a common, coherent terminology for specifications and to provide a concise reference of domain terms for the interested audience. For the dialog component, SWC [9] and SXCML (State Chart XML: State Machine Notation for Control Abstraction) [11] offer a language based on the State Machine concepts.

*B. User Stories*

User Stories in Software Engineering was first proposed by Cohn [4]. The author suggests to formalize User Stories in terms of artifacts for describing features and their acceptance criteria, with concrete examples about what should be tested to consider these features as "done". For that, the stories are formatted to fulfill two main goals: (i) assure testability and non-ambiguous descriptions and (ii) provide reuse of business scenarios. Figure 1 presents a template proposed by North [12] and Cohn [4].

```
Title (one line describing the story)
Narrative:
As a [role]
I want [feature]
So that [benefit]
Acceptance Criteria: (presented as Scenarios)
Scenario 1: Title
Given [context]
 And [some more context]...
When  [event]
Then  [outcome]
 And [another outcome]...
Scenario 2: ...
```

Figure 1.   Template for specifying User Stories, North [12] and Cohn [4].

According to this template, a User Story is described with a Title, a Narrative and a set of Scenarios representing the Acceptance Criteria. The Title provides a general description of the story, making reference to a feature that this story represents. The Narrative describes the referred feature in terms of the role that will benefit from the feature, the feature itself, and the benefit it will bring to the business. The Acceptance Criteria are defined through a set of Scenarios, each one with a Title and three main clauses: "*Given*" to provide the context in which the Scenario will be actioned, "*When*" to describe events that will trigger the Scenario and "*Then*" to present outcomes that might be checked to verify the proper behavior of the system. Each one of these clauses can include an "*And*" statement to provide multiple contexts, events and/or outcomes. Each statement in this representation is called Step.

In Behavior-Driven Development (BDD) [1], the user's point of view about the system is captured by the User Stories, described according to the template shown in Figure 1. BDD approach assumes that clients and teams can communicate using this semi-structured natural language description, in a non-ambiguous way (because it is supported by test cases).

Following this assumption, we have defined a conceptual model to represent user requirements. Our focus is on functional requirements. A functional requirement defines statements of services that the system should provide, how the system should react to particular inputs, and how the system should behave in particular situations. To assure that the system behaves properly, requirements should be expressed in a testable way. Figure 2 presents the conceptual model that explains how testable requirements are formalized by our approach.

As show in Figure 1, a requirement is expressed as a set of User Stories (US) encompassing a Narrative and a set of Acceptance Criteria. Acceptance Criteria are presented as Scenarios and are composed by at least three main Steps ("*Given*", "*When*" and "*Then*") that represent Behaviors which the system can answer. Behaviors handle actions on Interaction Elements in the User Interface (UI) and can also mention examples of data that are suitable for testing them. These concepts and rules are defined as classes and axioms in the proposed ontology presented hereafter.

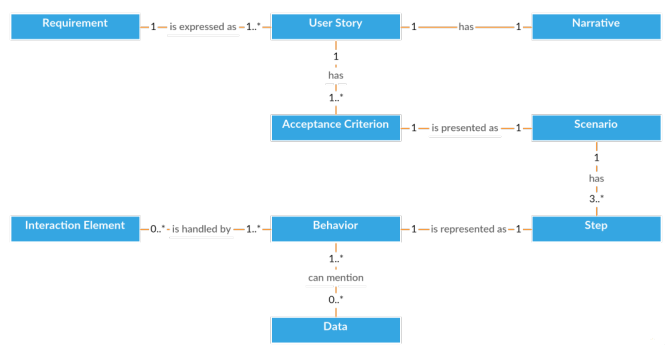

Figure 2.   Conceptual Model of User Requirements

## III. ONTOLOGY MODELING

The ontology we introduce in this paper borrows many concepts from languages and vocabularies in the literature. From Camaleon [6] and UsiXML [7] we get concepts of abstract and concrete UIs. Presentation and definition of graphical components comes from W3C MBUI [10]. The OWL specification of the ontology (W3C Web Ontology Language) encompasses concepts (behavior and presentation aspects) of graphical components commonly used to build web and mobile applications and it also contains textual representations that are used to describe how users interact with those graphical components. Previous works with SWC [8] also inspired the concepts used for describing the dialog.

The ontology has been modeled in Protégé 5.0. Figure 3 presents the root classes of the ontology. The classes *Dialog*, *Presentation* and *Platform* model the concept of a *Prototype*. A *Prototype* is built for at least one *Platform* and is specified by no more than one *Dialog* and one *Presentation*. A *Dialog* is performed by a *State Machine* (detailed in section 3C) and a *Presentation* is performed by the *Interaction Elements* (detailed in section 3A). Likewise, the classes *Narrative*, *Scenario*, *Step* and *Task* model the concept of a *User Story*. A *User Story* is described by exactly one *Narrative* and some *Scenarios*. A *Scenario* is an occurrence of only one *Task* and is a set of *Steps*. A *Step* shall represent some *Event*, *Condition* and/or *Action* that are *Transition* elements from the *State Machine*, performing the *Dialog* component of a *Prototype*.

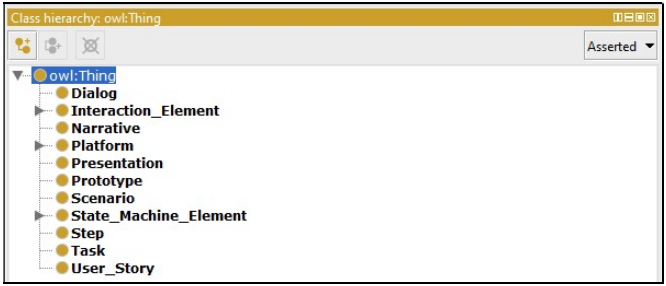

Figure 3. Root classes of the ontology

The current version of the ontology bears an amount of 422 axioms (being 277 logical axioms), 56 classes, 33 object properties, 17 data properties and 3 individuals. A visual representation of all the concepts can be found at https://goo.gl/IZqSJ0 and its complete specification in OWL can be found at https://goo.gl/1pUMqp.

### A. Modeling of Interaction Elements

Interaction Elements in the ontology represent an abstraction of GUI components in web and mobile platforms. Figure 4 illustrates a hierarchy of Interaction Elements. The four main superclasses are Container, Information Component, Input Control and Navigational Component. The first one contains elements that group other elements in a User Interface, such as Windows and Field Sets. The second one contains elements in charge of displaying information to the users such as Labels and Message Boxes. The third one represents elements in which users provide inputs to the system such as Buttons and Text Fields. Finally, the last one contains elements useful to navigate through the system such as Links and Menus. Some elements like Dialog Windows, for example, are inherited by more than one superclass, once they keep semantic characteristics of Containers and Information Components as well.

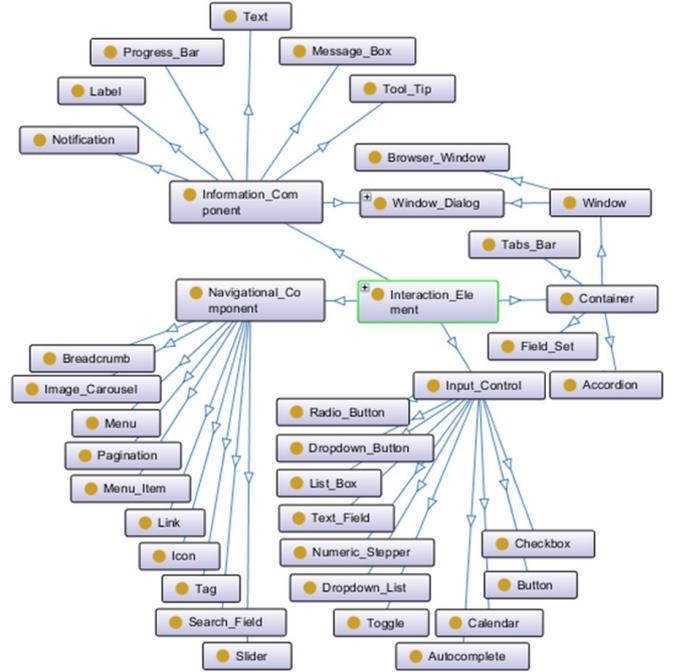

Figure 4. Cloud of User Interface Elements

### B. Data Properties

Data Properties have been created to semantically describe data domains used by each Interaction Element. The root tree shown in Figure 5a gives an overview of the properties created, while Figure 5b expands the Data Property "*message*", showing that this kind of data is used by the Interaction Elements "*Message Box*", "*Notification*", "*Tool Tip*" and "*Modal Window*". "*Message*" has also been defined to range the primitive data *String*.

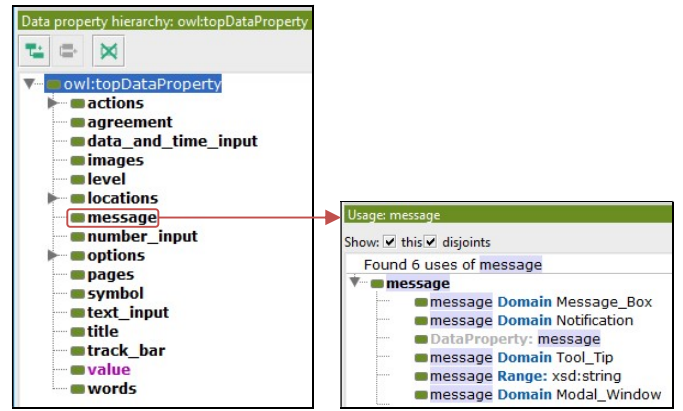

Figure 5. **(a)** Left: Data Properties. **(b)** Right: Data Property "*message*"

## C. Modeling of the State Machine

The dialog part of a User Interface is described in the ontology using concepts borrowed from abstract State Machines as illustrated by Figure 6. A Scenario meant to be run in a given UI is represented as a Transition. States are used to represent the original and resulting UIs after a transition occur. Scenarios in the Transition state always have at least one or more Conditions (represented in Scenarios by the "*Given*" clause), one or more Events (represented in Scenarios by the "*When*" clause), and one or more Actions (represented in Scenarios by the "*Then*" clause). The clauses "*Given*", "*When*" and "*Then*" have been modeled as Individuals of each respective class.

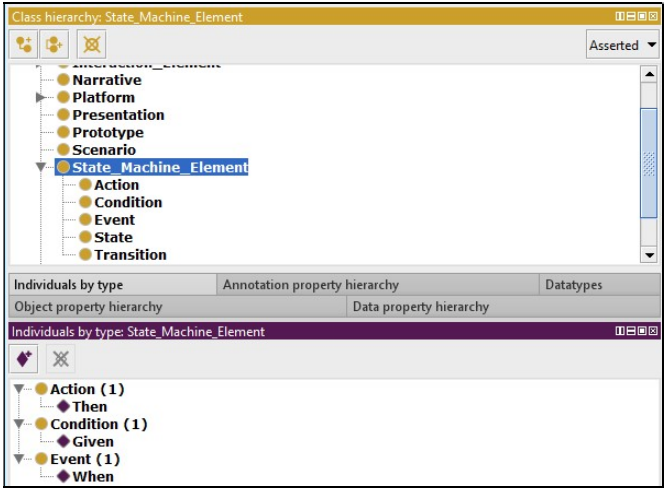

Figure 6. State Machine Elements and their Individuals

## D. Object Properties

Relationships between individuals in classes are represented as Object Properties. We have classified those properties in *Relations* and *Behaviors*. *Relations* group conceptual relationships between objects from internal classes, i.e. objects that do not directly address user's interaction. *Behaviors* group conceptual relationships between user's interactions and Interaction Elements on the UI. These two groups are explained hereafter.

### 1) Relations:

Figure 7 presents the set of main relationships between objects of internal classes defined in the ontology. As an example, Figure 8 presents the relations between elements in the State Machine. As a sub property of *Relations*, objects from the *Dialog* class are composed by some *States* and *Transitions*. This relationship is described by the property *isComposedBy* (left-side of Figure 8). Accordingly, objects from the *Transition* class are triggered by a sequence of some *Conditions*, *Events* and *Actions*. This relationship is described by the property *isTriggeredBy* (right-side of Figure 8).

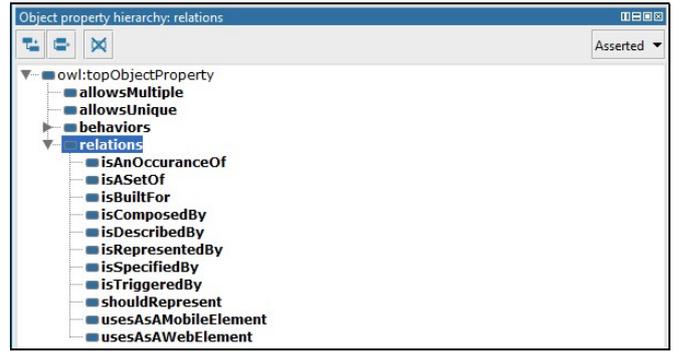

Figure 7. Object Property "*Relations*"

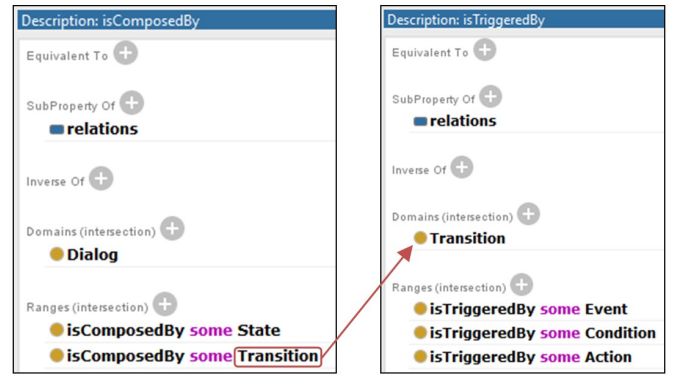

Figure 8. Object Properties *isComposedBy* (**left**) and *isTriggeredBy* (**right**)

### 2) Behaviors:

Following our ontological approach, the concept *Behaviors* is used to describe how users are supposed to interact with the systems whilst manipulating graphical elements of the User Interface. At first, behaviors can be described in a kind of structured natural language so that they can also be read by humans. As illustrated by Figure 9, the specification of behaviors encompasses when the interaction can be performed (using *Given*, *When* and/or *Then* clauses) and which graphical elements (i.e. *Radio Button*, *CheckBox*, *Calendar*, *Link*, etc.) are suitable to implement the expected system behavior. In the example below, the behavior receives two parameters: a "*$elementName*" and a "*$locatorParameters*". The first parameter is associated to data, the second parameter refers to the Interaction Element supported by this behavior: "*Radio Button*", "*CheckBox*", "*Calendar*" and "*Link*". To comply with semantic rules, the behavior "*I chose \"$elementName\" referring to \"$locatorParameters\"*" shown in Figure 9 can be modelled into a predefined behavior "*chooseReferringTo*" as shown in Figure 10.

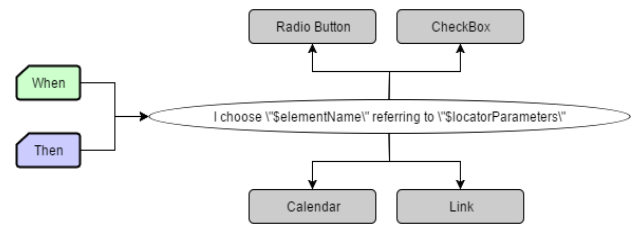

Figure 9. Components on the ontology used to specify a behavior

Figure 10. Behavior "*chooseReferringTo*".

Our ontology includes a large set of predefined behaviors. Some of them are show at Table 1. Notice that each *Behavior* is associated to diverse transition components (*Context*, *Event* and/or *Action*) that compose a *Transition*. The column *Interaction Elements* in Table 1 enlists the set of Interaction Elements that can fit to trigger a particular behavior.

TABLE I. SOME OF PREDEFINED BEHAVIORS ON THE ONTOLOGY

| Behavior | Transition | | | Interaction Elements |
|---|---|---|---|---|
| | C | E | A | |
| *choose* | | | | Calendar<br>Checkbox<br>Radio_Button<br>Link |
| *chooseByIndexInTheField* | | | | Dropdown_List |
| *chooseReferringTo* | | | | Calendar<br>Checkbox<br>Radio_Button<br>Link |
| *chooseTheOptionOfValueInTheField* | | | | Dropdown_List |
| *clickOn* | | | | Menu<br>Menu_Item<br>Button<br>Link |
| *clickOnReferringTo* | | | | Menu<br>Menu_Item<br>Button<br>Link |
| *doNotTypeAnyValueToTheField* ≡ *resetTheValueOfTheField* | | | | Text_Field |
| *goTo* | | | | Browser_Window |
| *isDisplayed* | | | | Window |
| *setInTheField* ≡ *tryToSetInTheField* | | | | Dropdown_List<br>Text_Field<br>Autocomplete<br>Calendar |
| *typeAndChooseInTheField* | | | | Autocomplete |
| *willBeDisplayed* | | | | Text |

The vocabulary chosen to express each behavior has emerged from Scenarios specified in our past projects. It outlines only one of the several possible vocabularies to represent the same user's behaviors, and could be extended in the future by more representative phrases or expressions. Some synonyms concerning the user's goal have been also identified in order to increase the expressivity of the ontology. For example, the behavior `doNotTypeAnyValueToTheField` is considered equivalent to the behavior `resetTheValueOfTheField` as they perform or assert exactly the same action on the affected UI element, looking for the same output. Likewise, the behavior `setInTheField` is equivalent to the behavior `tryToSetInTheField` as they intend the same action, being the second one more suitable to express attempts of violation in the business rules for testing purposes.

## IV. VALIDATION

The ontology has been validated in two steps: at first, consistency has been continuously checked through the use of reasoners. Then, using a tool support, we applied the approach to a case study in the flight tickets e-commerce domain. We have developed tools for dealing with tests over Prototypes and for testing the implementation. A first tool named PANDA (Prototyping using Annotation and Decision Analysis) [13] was built to design and test UI Prototypes. A second tool was developed to testing Web Final UIs derived from the Prototypes. Both tools have been used to conduct the case study.

### A. Consistency Checking

For checking the consistency of the ontology we use the reasoners FaCT++, ELK, HermiT and Pellet. FaCT++ started identifying no support for the datatypes `xsd:base64Binary` and `xsd:hexBinary` used to range *images* and *symbols* in the Data Properties. Those properties have been used to define domains for objects in the classes *Image Carousel* and *Icon*, respectively. ELK has failed by no support to Data Property Domains as well as Data and Object Property Ranges. HermiT and Pellet have succeeded processing the ontology respectively in 4926 and 64 milliseconds, as presented in Figure 11.

Figure 11. Results of ontology processing: HermiT (top) and Pellet (bottom).

### B. Validation by a Case Study

To illustrate how the ontology can be used to support the specification of requirements and the testing automation for interactive systems, we have chosen a flight tickets e-commerce application. Figure 12 describes one of the User Stories from this case study with a Scenario for searching flights. Therein, the user should provide at least: the type of sought ticket (one-way or round trip), the departure and the arrival airports, the number of passengers, and finally the dates. In the Scenario "*One-Way Tickets Search*", a typical search of tickets is presented concerning a one-way trip from Paris to Dallas for 2 passengers on 12/15/2016. According to the business rule, the expected result for this search is a new screen presenting the title "*Choose Flights*", in which the user might select the desired flight from a list of flights matching his search.

```
User Story: Flight Tickets Search
Narrative:
As a frequent traveler
I want to be able to search tickets, providing locations and
dates
So that I can obtain information about rates and times of
the flights.
Scenario: One-Way Tickets Search
Given I go to "Find flights"
When I choose "One way"
And I type "Paris" and choose "CDG - Paris Ch De Gaulle,
France" in the field "From"
And I type "Dallas" and choose "DFW - Dallas Fort Worth
International, TX" in the field "To"
And I choose the option of value "2" in the field "Number of
passengers
And I choose "12/15/2016" referring to "Depart"
And I click on "Search"
Then will be displayed "Choose Flights"
```

Figure 12. User Story for *Flight Ticket Search* in the testing template format

*1) Ontology Support for Testing Prototypes using PANDA:*

PANDA [13] is a tool support for the creation and testing of UI Prototypes built upon an ontology. PANDA starts by reading an OWL file describing our ontology. Using the inner organization of ontological classes, PANDA dynamically instantiates a palette of widgets (see Figure 13) that can be used to build a Prototype. From an interaction point of view, the construction of Prototypes is done by performing drag and drop operations. From a storage point of view, a Prototype is an XML file that describes a composition of widgets whose description is semantically annotated by elements of our ontology.

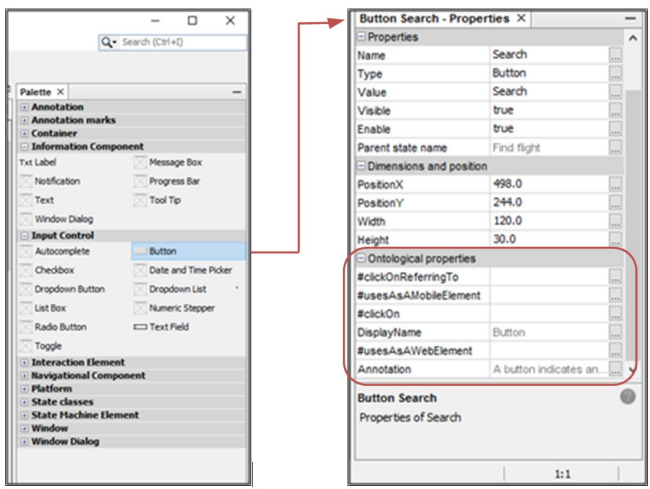

Figure 13. Pallets with the widget Button and its properties extracted from the ontology.

For the construction of the palette, PANDA uses a description of a widget we called "*OntologicalClass*" which feature its name, list of subclasses and set of properties. This ontological class has been defined as a generic class that is customized through its properties. Indeed, those classes represent each component of a Prototype in PANDA and its behaviors regarding their usage in the prototyping tool: they are placed in an edition area in which the user can edit the instance of a property. Thus, for the Presentation component, PANDA uses a flexible structure that allows to dynamically instantiate the set of widgets that will be used to build Prototypes.

PANDA creates a category for each superclass including: Container, Information Component, Input Control, Interaction Element, Navigational Component, Platform, State Machine Element, Window and Window Dialog. Each category contains a set of widgets defined by the classes inheriting the superclass. As for the properties, ontological classes are displayed in the property window in the category "Ontological properties". Each property identified in the ontology is therefore inserted in the list of properties of the class with a name and a value.

For the Dialog component, our ontology encompasses behavioral properties to describe the interaction supported by a class. For example, a Button must feature a behavioral property "*clickOn*" which indicates that buttons support an event click. Click events allow the designer to specify interactions on widgets. If a button has a behavioral property "*clickOn*", PANDA adds an event handler to handle click events when users interact with the Prototype.

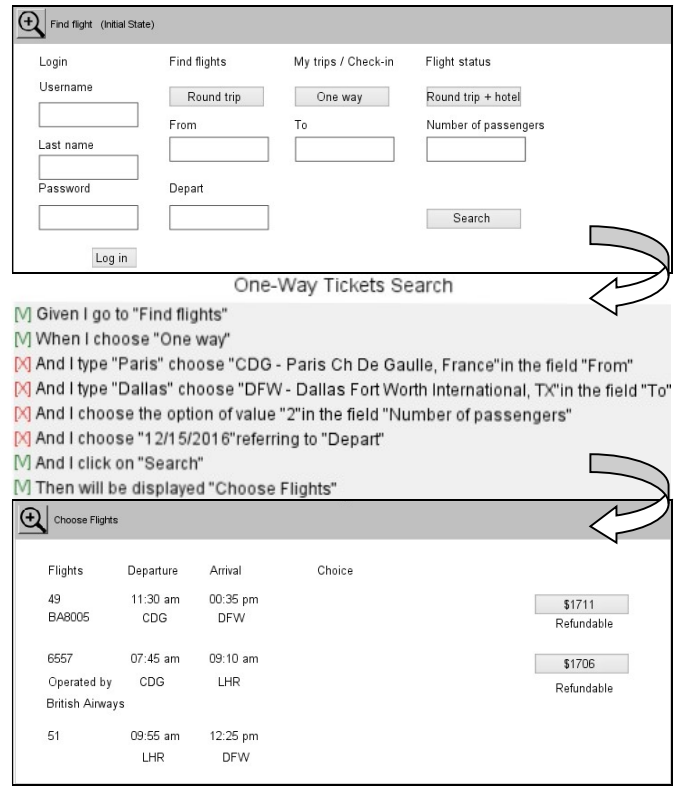

Figure 14. A State Machine Transition between sketches of a PANDA Prototype for the User Story "Flight Tickets Search". From top to bottom: the initial State "Find Flights", a Transition represented by the Scenario "One-Way Tickets Search", and finally the resultant State "Choose Flights".

Figure 14 shows how Scenarios are tested in PANDA. For each Step of Scenarios, PANDA assess actions with respect to widget properties defined in the ontology. For example, in the Step "And I click on 'Search'", PANDA looks for any widget named "Search" in the initial State, and check if the description of the widget in the ontology support the behavior "clickOn". The results of the tests are displayed by a colored symbol next to each Step, a red "X" representing failure, a green "V" representing success, and a black "?" representing an untested Step.

*2) Ontology Support for Testing Web Final UIs:*

To test the Scenarios over Web Final UIs, we have employed a set of frameworks to provide automated simulation of user's interaction. More specifically, we have used Selenium WebDriver to run navigational behavior as well as JBehave and Demoiselle Behave to parse Scenario scripts. The ontology is charged as a *CommonSteps* Java Class, pre-defining behaviors that can be used when writing Scenarios, and where each action and/or assert for each behavior is defined. This class implements the dialog component and contains all the knowledge about how performing the mentioned behaviors on the UI elements, thus when using them to write Scenarios, tests are delivered without any additional effort of implementation. Hence, methods in this class have been written for every Step addressed on the ontology. As illustrated in Figure 15, behaviors "*When/Then I choose '...' referring to '...'*" are addressed to the Selenium method *click()*, with the appropriated sequence of actions to perform this task on the Final UI. As this behavior can be performed only in *Radio Buttons*, *Check Boxes*, *Links* or *Calendars*, the concrete instance of any of these elements are searched on the Presentation layer.

```
@When(value = "I choose \"$elementName\" referring to \"$locatorParameters\"", priority = 10)
@Then(value = "I choose \"$elementName\" referring to \"$locatorParameters\"", priority = 10)
public void informWithParameters(String elementName, List<String> locatorParameters) {
    locatorParameters = DataProviderUtil.replaceDataProvider(locatorParameters);
    Element element = runner.getElement(currentPageName, elementName);
    element.setLocatorParameters(locatorParameters);
    if (element instanceof Radio) {
        ((Radio) element).click();
    } else if (element instanceof CheckBox) {
        ((CheckBox) element).click();
    } else if (element instanceof Link) {
        ((Link) element).click();
    } else if (element instanceof Calendar) {
        ((Calendar) element).click();
    } else {
        throw new BehaveException(message.getString("exception-invalid-type", element.getClass().getName()));
    }
}
```

Figure 15. Behavior "*chooseRefferingTo*" being structured as a Java method

The Presentation component includes the *MyPages* Java Class that makes the mapping between abstract UI elements of the ontology and the concrete/final UI components instantiated on the interface being tested. For that purpose, we make use of annotations in Java code following the Page Objects pattern [14] as illustrated in Figure 16. UI components are identified through their XPath references or some other unique ID eventually used for some frameworks to implement the interface. This link is essential to allow the framework to automatically run the Steps on the right components on the Final UI.

```
public class MyPages {
    @ScreenMap(name = "Find Flights", location = "..")
    public class MainPage {
        @ElementMap(name = "Search", locatorType = ElementLocatorType.XPath, locator = "…") // concrete UI component
        private Button Search; // abstract UI element
        …
    }
}
```

Figure 16. Concrete and Abstract UI elements being associated in a Java class

For behaviors not addressed by the ontology, the *MySteps* Java Class allows developers and testers to set their own business behaviors and implement as well how they should be attended by the Selenium methods on the UI components. For both classes the main incomes are behaviors extracted from the User Stories that can be represented in simple packages of text files.

In short, once the ontology is charged, it is enough to identify on the Final UI under testing the concrete UI elements that were instantiated to represent abstract UI elements. Afterwards, when Scenarios are triggered, the application runs and Selenium performs Step by Step the specified behaviors, reporting testing results either by the JUnit green/red bar or by JBehave reports with the context and attached print-screens of each identified failure.

*C. Discussion*

The ontology presented in this paper only describes behaviors that report Steps of Scenarios performing actions directly on the User Interface through Interaction Elements. Thus, the ontological model is domain-free, which means that it is not dependent of business characteristics that are described in the User Stories. Specific business behaviors shall be specified only for the systems they make reference, not affecting the whole ontology. Therefore, it is possible to reuse Steps in multiple testing Scenarios. For example, the ontological behaviors *goTo*, *choose*, *typeAndChooseInTheField*, *chooseTheOptionOfValueInTheField*, *chooseReferringTo*, *clickOn* and *willBeDisplayed* presented in Figure 12 can be easily all reused for any other Scenario of any other system requiring those kind of user's actions.

However, it brings the need to specify Scenarios in the user interaction level, writing Steps for each click, selection, typing, etc. A possible solution to avoid it would be to work with higher level behaviors that are described by user's tasks. Nonetheless, user's tasks often contain information from specific application domains. For example, high level Steps like "*When I search for flights to 'Destination'*" encapsulate all low level behaviors making reference for individual clicks, selections, etc., however it also contains information that refers to the airline domain (i.e. behavior "*search for flights*"). So that Step would only makes sense on that particular application domain. For further researches, it could be interesting to investigate domain ontologies to be used in parallel with our ontology, defining a higher level business vocabulary database in which business behaviors could be mapped to a set of interaction behaviors, covering recurrent Scenarios for a specific domain, and avoiding them to be written every time a new interaction may be tested.

Another aspect to be discussed is that even having mapped synonyms for some specific behaviors, our approach does not provide any kind of semantic interpretation, i.e. the Steps might be specified exactly as they were defined on the ontology. The use of the JBehave plugin for Eclipse has helped us to know visually (through different colors) on real time if some Step that is being written exists or not on the ontology. This resource reduces the workload to remember as exactly some behavior has been described on the ontology.

At first glance nonetheless the restricted vocabulary seems to bring less flexibility to designers, testers and requirements engineers, but on the other hand, it establishes a common vocabulary, avoiding typical problems of ambiguity and incompleteness in requirements and testing specifications. Naturally, investigating the use of Natural Language Processing (NLP) techniques could improve the specification process, adding more flexibility to write Scenarios that could be in some extent semantically interpreted to meet the behaviors described on the ontology. This issue is certainly a worthwhile topic for further researches.

It is also worthy of mention that the concepts and definitions in the ontology presented herein are only one of the possible solutions for addressing and describing behaviors and their relations with UIs. Despite being based on well-known languages such as MBUI, UsiXML and SCXML besides being provided ready to use for a new development project, we consider that the ontology might evolve to include other behaviors, concepts and relationships that could be eventually more representative for some contexts of development. It could be made through the use of direct imports in the ontology or simply adding new more expressive behaviors to the Object Property "behaviors" and linking them to the appropriate set of Interactive Elements.

Finally, when representing the various Interaction Elements that can attend a given behavior, the ontology also allows extending multiple design solutions for the UI, representing exactly the same requirement in different perspectives. Thus even if a Dropdown List has been chosen to attend for example a behavior *setInTheField* in a Prototype, an Auto Complete field could be chosen to attend this behavior on the Final UI, once both UI elements share the same ontological property for this behavior under testing. This kind of flexibility makes tests pass, leaving the designer free for choosing the best solutions in a given time of the project, without modifying the behavior specified for the system.

## V. CONCLUSION

In this paper we have presented a behavior-based ontology aiming at test automation that can help to validate functional requirements when building interactive systems. The proposed ontology acts as a base of common vocabulary articulated to map user's behaviors to Interaction Elements in the UI which allows us to automate tests. The ontology also provides important improvements in the way teams should write requirements for testing purposes. Once described in the ontology, behaviors can be freely reused to write new Scenarios in natural language, providing test automation with little effort from the development team. Moreover, it allows specifying tests in a generic way that can be reused along the development process. For that reason, we are also investigating the use of the ontology to test model-based artifacts such as low-fidelity Prototypes and Task Models. Tests in these artifacts could be conducted through a static verification of their source codes and would help to integrate testing in a wider spectrum of artifacts commonly used to build interactive systems.

We have also presented tools that demonstrate how this ontology can support testing of interactive systems. So far, only interactive Prototypes built in PANDA can be tested by the ontology once it requires that tools are able to read and support the set of described behaviors. On the other hand, tests in Web Final UIs can run independently of the frameworks used to build these UIs. It is possible because tests provided by our tool assess the concrete UI elements found on the interface in the final HTML page.

### A. Future Works

We envision a set of future works in this theme including experiments to evaluate the acceptance of the approach in a case study with users, especially when manipulating more complex behaviors in real cases of software development. It would be useful to collect data about the effectiveness and the workload when specifying tests using the ontology. Other case studies including mobile platforms are planned as well.

Future discussions might also consider having ontologies as knowledge bases, keeping specific behaviors for specific groups of business models in domain ontologies. It would allow us to also reuse entire business Scenarios in systems sharing similar business models.